\documentclass[journal=jacsat,manuscript=article]{achemso}

\usepackage[version=3]{mhchem} 

\author{Sk Mujaffar Hossain}
\affiliation[IKST]
{Indo-Korea Science and Technology Center (IKST), Bangalore, India}
\author{Dobin Kim}
\affiliation[KIST2]
{KHU-KIST Department of Converging Science and Technology, Kyung Hee University, Seoul 02447, Korea}
\author{Jaehyun Park}
\affiliation[KIST1]
{Extreme Materials Research Center, KIST, Seoul 02792, Korea}
\author{Seung-Cheol Lee}
\email{leesc@kist.re.kr}
\affiliation[IKST]
{Indo-Korea Science and Technology Center (IKST), Bangalore, India}
\alsoaffiliation[EMRC KIST]
{Electronic Materials Research Center, Korea Institute of Science $\&$ Technology, Korea}

\author{Satadeep Bhattacharjee}
\email{s.bhattacharjee@ikst.res.in}
\affiliation[IKST]
{Indo-Korea Science and Technology Center (IKST), Bangalore, India}

\title[An \textsf{achemso} demo]
  {Tunable Thermal Expansion in Functionalized 2D Boron Nitride: A First-Principles Investigation}

\abbreviations{IR,NMR,UV}

\begin{document}

\begin{tocentry}

	\includegraphics[width=5.6cm]{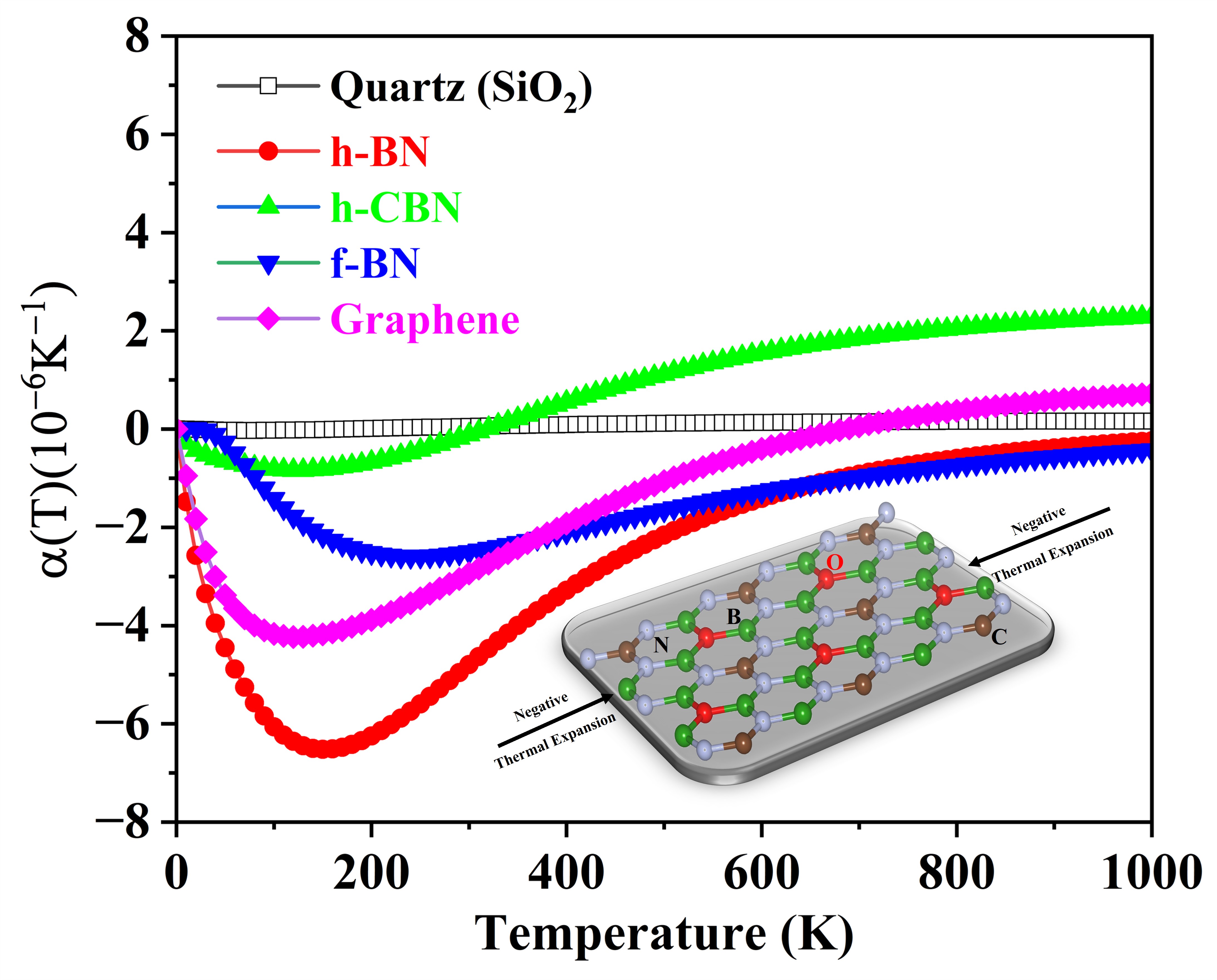}
\end{tocentry}

\begin{abstract}
 This study investigates the thermal expansion coefficient of two-dimensional (2D) functionalized boron nitride (f-BN) materials using first-principles density functional theory (DFT). Two-dimensional materials, particularly hexagonal boron nitride (h-BN), have attracted significant attention due to their exceptional mechanical, thermal, and electronic properties. However, the influence of functionalization on the thermal expansion behavior remains largely unexplored. In this work, DFT calculations are employed to analyze how different functionalized forms of h-BN impact the thermal expansion of BN sheets. Density functional perturbation theory (DFPT) and the quasiharmonic approximation (QAH) are utilized to determine the thermal expansion coefficient over a range of temperatures. The results reveal that functionalization induces notable modifications in the in-plane thermal expansion of BN, affecting material stability and suggesting potential applications in nanoelectronics and thermal management. This investigation provides critical insights into the tunability of the thermal properties of 2D BN, underscoring its suitability for next-generation flexible and high-performance devices.
\end{abstract}

\section{Introduction}
New strategies always pave the way for building and breeding materials with exciting properties and multiple applications by creating functionalized materials using existing parent materials of all dimensions (0 to 3D). Various functionalized materials have been proposed and discovered, leaving their mark in many fields of science and technology. Since the beginning of material discovery, the application of functionalized materials has covered all areas, such as solar cells, catalysis, energy storage, superconductivity, artificial intelligence, thermal energy, and more. Many 2D functional materials, including graphene, TMDs, phosphorene, MXene, and h-BN, have been explored in numerous fields by the research and scientific community. It is challenging to discuss every application of these 2D materials as each material performs excellently in different areas. Here, our work focuses on one of the exciting properties of 2D materials: thermal expansion, which can be zero, positive, or negative.

Thermal expansion in crystal materials refers to the tendency of the crystal lattice to change its dimensions with temperature fluctuations. This phenomenon occurs because the atomic vibrations within the crystal increase with temperature, causing the atoms to occupy more space and thus expanding the crystal. The degree of thermal expansion varies among different materials and is quantified by the linear thermal expansion coefficient ($\alpha$). This property is crucial in applications where precise dimensional stability is required, such as in aerospace engineering, electronics, and precision instruments, as it helps to predict how materials will behave under thermal stress. Understanding and controlling thermal expansion is essential for developing reliable and efficient materials for high-performance applications.

Thermal expansion in 2D materials is a critical property that influences their performance in various applications \cite{zhong2022unified}. Unlike bulk materials, 2D materials  
exhibit unique thermal expansion behaviors due to their atomic thickness \cite{hu2018mapping}. When heated, the atoms in these materials vibrate more intensely, causing the material to expand \cite{cai2019high}. However, the degree of expansion can vary significantly depending on the material and its environment \cite{zhong2022unified}.

One study by \citeauthor{hu2018mapping}\cite{hu2018mapping} introduced a novel approach to measure the thermal expansion coefficients (TECs) of freestanding 2D materials at the nanometer scale using scanning transmission electron microscopy combined with electron energy-loss spectroscopy \cite{hu2018mapping}. This method allowed for precise mapping of TECs in materials like graphene, MoS$_2$, MoSe$_2$, WS$_2$, and WSe$_2$ \cite{hu2018mapping}. The results showed that 2D materials generally have lower TECs compared to their bulk counterparts, which is beneficial for applications requiring high dimensional stability\cite{zhong2022unified}. Another study by \citeauthor{zhong2022unified}\cite{zhong2022unified} demonstrated a technique to measure the TECs of 2D materials by tracking atomic vibrations using laser light \cite{zhong2022unified}. This method confirmed that the TECs of 2D materials fall within a narrower range than previously thought, providing engineers with more flexibility in designing devices\cite{zhong2022unified}. Understanding and controlling the thermal expansion of 2D materials is essential for developing reliable and efficient electronic devices, as mismatches in thermal expansion between the 2D material and its substrate can lead to thermal stress and device failure \cite{zhong2022unified}. By accurately measuring and predicting the thermal expansion properties, researchers can design better materials for applications in nanoelectronics, optoelectronics, and other high-performance devices\cite{zhong2022unified,hu2018mapping}.

Negative thermal expansion (NTE) in 2D materials is a fascinating phenomenon where materials contract upon heating, contrary to most materials that expand \cite{liang2021negative}. This unusual behavior arises from specific physical processes, such as transverse vibrational modes, rigid unit modes, and phase transitions \cite{liu2011origin,liu2014thermal,cabras2019micro}. This property is significant because it allows for the design of composites and devices that can maintain dimensional stability across a range of temperatures \cite{dubey2024negative}. For instance, NTE materials can be used as thermal expansion compensators in electronics and aerospace components, where precise dimensional control is crucial\cite{dubey2024negative}. Additionally, 2D materials with NTE properties can be integrated into nanoscale devices, ensuring they operate without expansion or contraction over the required temperature range\cite{sarikurt2022negative}. This capability opens up new possibilities for advanced material design and innovative applications in various high-tech fields\cite{dubey2024negative,sarikurt2022negative}. Interfaces is everywhere and it also play an important role in many device application, since the effect of NTE at the interface is also crucial. NTE helps in maintaining dimensional stability by counteracting the expansion of the substrate, which is particularly important in microelectronics and aerospace components\cite{xiao2024effects}. This property minimizes thermal stress and strain, reducing the risk of mechanical failure and improving the longevity of the device\cite{xiao2024effects}. Additionally, NTE at the interface can improve thermal management by ensuring consistent thermal conductivity and reducing interfacial thermal resistance, leading to more efficient heat dissipation\cite{xiao2024effects,yao2024hydrogen}.

Here using DFT, our study focused on the 2D h-BN and its functional BN (f-BN) material for exploring their NTE behavior. For pure h-BN the coefficient of thermal expansion has been reported by \citeauthor{sevik2014assessment}\cite{sevik2014assessment} and \citeauthor{olaniyan2021first}\cite{olaniyan2021first} where they have theoretically shown the negative thermal expansion coefficient of -6.6 X 10$^{-6}K^{-1}$ using DFTP and QHA in the temperature interval of 0 to 300K. The h-BN
has a typical layered structure like graphite with two dimensional (2D) layer, alternating B and N atoms are linked with each other via strong B–N covalent bonds; whereas the 2D layers are held together by weak van der Waals forces. Unlike the case of graphite, the interlayer stacking pattern in the h-BN
features its B atoms in every consecutive BN layer siting exactly above or below the N atoms in the adjacent layers. Such structural characteristics imply the polarity of B–N bonds, i.e. the partially ionic character of the covalent B–N bonds. Electron pairs in sp$^2$-hybridized B–N $\sigma$ bonds are more confined to the N
atoms due to their higher electronegativity; and the lone pair of electrons in the N p$_z$ orbital is only partially delocalized with the B p$_z$ orbital, in contrast to the equally contributed and evenly distributed electrons along the C–C bonds of graphite layers. The unique structural features of h-BN endow itmany other important electrical, optical, and chemical characteristics. For example, the reduced electron-delocalization in the BN $\pi$ bonds causes a large band gap and leads to the electrically insulating nature and colourless appearance of the material. These properties make h-BN very useful as insulating and thermally conductive fillers, deep ultraviolet light sources, dielectric layers, cosmetic products, microwave-transparent shields, etc\cite{jiang2015recent}. Furthermore, h-BN is highly thermally and chemically stable, and thus is also widely used for durable high-temperature crucibles, anti-oxidation lubricants, protective coatings, etc. in industry\cite{paine1990synthetic}.

It is expected that many novel properties can emerge from a material through its smart functionalization, either physical or chemical. This is also proven to be applicable in case of h-BN. Its properties can be tailored, and many brand-new features and applications can be created directly via such functionalization. Since the B–N bonds in an h-BN structure have a partial ionicity, its B and N atoms are partially positively (electron deficient centers) and negatively (electron rich centers) charged, respectively. This property makes the B sites attackable by nucleophilic groups, while the N sites are reactive with electrophilic ones. Heteroatoms C and O could be experimentally (like CVD\cite{chen2021growth} and PLD\cite{godbole2024light}) introduced into BN sheets through chemical functionalization \cite{weng2016functionalized}. We construct the f-BN and performed comparison study of free standing f-BN with h-BN and h-CBN with their coefficient of thermal expansion and other electronic properties.

\section{Computational Details}
The equilibrium in-plane lattice parameters at any temperature T are determined by fitting the Brich-Murnaghan equation of states\cite{sevik2014assessment,birch1947finite} to Helmholtz free energy calculated by the following equations
\begin{equation}
    F(a_i, T) = E(a_i) + \int[\frac{h\omega^i}{2} + \kappa_BT\int ln(1 - exp(-\frac{h\omega^i}{\kappa_BT}))]\rho(\omega^i)d\omega^i
\end{equation}

 During the fitting procedure, twelve different lattice parameters, \(a_i\), around first-principles equilibrium lattice constant are considered (between 0.995 to 1.006\(a_0\) for h-BN\cite{sevik2014assessment,demiroglu2021extraordinary}, h-CBN\cite{olaniyan2021first} and f-BN, 
 For each structure with lattice constant \(a_i\), the exact vibrational frequencies are obtained using the PHONOPY code\cite{togo2013phonopy} which can directly use the force constants calculated by DFPT as implemented in the VASP package\cite{kresse1993ab,kresse1996efficiency,kresse1996software,kresse1994norm,kresse1999ultrasoft}. Thermal properties were computed using quasiharmonic approximation (QHA)\cite{abraham2018thermal,togo2013phonopy,demiroglu2021extraordinary,pavone1993ab} implemented in PHONOPY\cite{togo2013phonopy}. The coefficient of thermal expansion (CTE), \(\alpha(T)\) was calculated using the following equation\cite{wang2017mechanical}:

\begin{equation}
    \alpha(T) = \frac{1}{V(T)_0}\frac{dV(T)_0}{dT}
\end{equation}

 The coefficient of Linear thermal expansion \(\alpha_L\) is derive using \(\alpha_L = \frac{\alpha(T)}{2}\) for 2D sheets such as monolayer graphene, TMD, h-BN etc,  and  \(\alpha_L = \frac{\alpha(T)}{3}\) for 3D bulk systems\cite{slack1975thermal,wang2017mechanical,brito2020thermodynamic}

 The ab-initio calculations are performed using the VASP code based on density functional theory\cite{martin2020electronic}. The projector augmented pseudopotentials (PAW)\cite{blochl1994projector,kresse1999ultrasoft} from the standard distribution are incorporated in the calculations. The generalized gradient approximation\cite{perdew1996generalized} (GGA) functional is used for electronic exchange correlations in its PBE parameterization. The thermal properties are strongly influenced by simulation parameters like wave cutoff energy, Brillouin zone sampling, and supercell size \cite{sevik2014assessment}, since we adopted the following parameters: 750 eV plane wave cutoff energy for all the system, 5x5x1 for h-BN and 4x3x1 for h-CBN, 
 as supercell structure, and 3x3x1 for h-BN, 8x3x1 for h-CBN, 
  as k-point mesh for the Brillouin zone sampling of the supercell structure. To compute the LTEC we kept the vacuum spacing at 20Å and varied the in-plane lattice constant within 0.5\% for 12 different slab volumes. For determine the electronic bandgap we have used HSE06\cite{heyd2003hybrid} hybrid functional.

\subsection{Functional Boron Nitride (f-BN)}
\begin{figure}[htp]
    \centering
    \includegraphics[width=15cm]{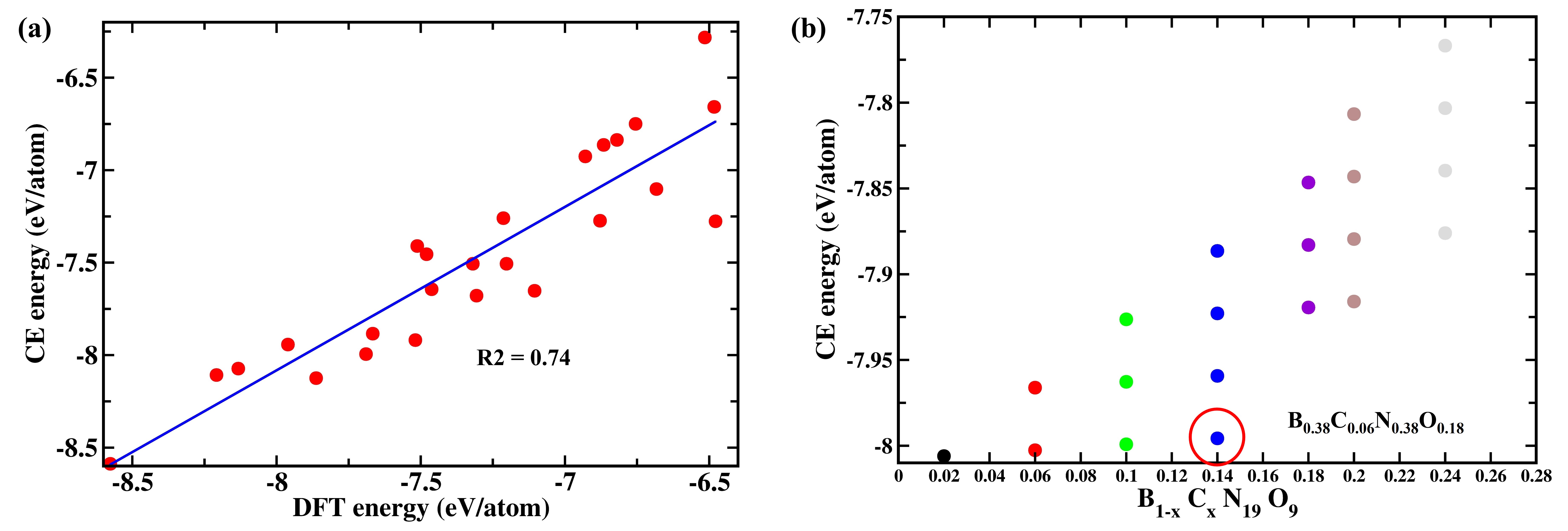}
    \caption{(a) Cluster expansion (CE) model and DFT recorded data are clustered around $(y = x)$ line. (b) CE energy of different composition of f-BN}
    \label{fig:fig_0}
\end{figure}

To build simplistic functional boron nitride (f-BN) we have considering 2x2x1 supercell of pure h-BN which has 4 B and 4 N atoms in the cell. We replace one B and N atom by C and O atom to construct the f-BN and the system crystal compositions become B$_3$N$_3$CO (12.5\% C and O). We optimized the f-BN structure by considering different combinations of replacing B and N by C and O and the value of formation energy help obtained most favorable structure as shown in Figure 1. This f-BN is a simplistic computational model of order crystal structure but experimentally the composition could be different and we tried to realized one such f-BN structure of B 38.0\%, N 38.0\% C 0.06\%, and O 18.0\% atomic compositions. To avail of this composition, we have performed cluster expansion\cite{chang2019clease} (CE) calculation and build an ECI model by considering small supercells like 2x2x1, 3x1x1, and 4x1x1 respectively. This model (Figure \ref{fig:fig_0}a) helps us design the closest composition of the functional boron nitride or f-BN, we follow the \citeauthor{das2023computational}\cite{das2023computational} reference for CE calculation details. The ECI model fit between CE and DFT energy with $(y = x)$ line (Figure \ref{fig:fig_0}a) and the fitting is reasonably good with an R$^2$ value of 0.74. The red circle represents the stable composition concerning the CE energy (Figure \ref{fig:fig_0}b). The stable f-BN system has 50 atoms in the system and to reduce the computational efforts here we performed frozen phonon calculation to calculate the LTEC.

\section{Results and discussion}

 \subsection{Structural and Electronic Properties}
\begin{figure}[htp]
    \centering
    \includegraphics[width=15cm]{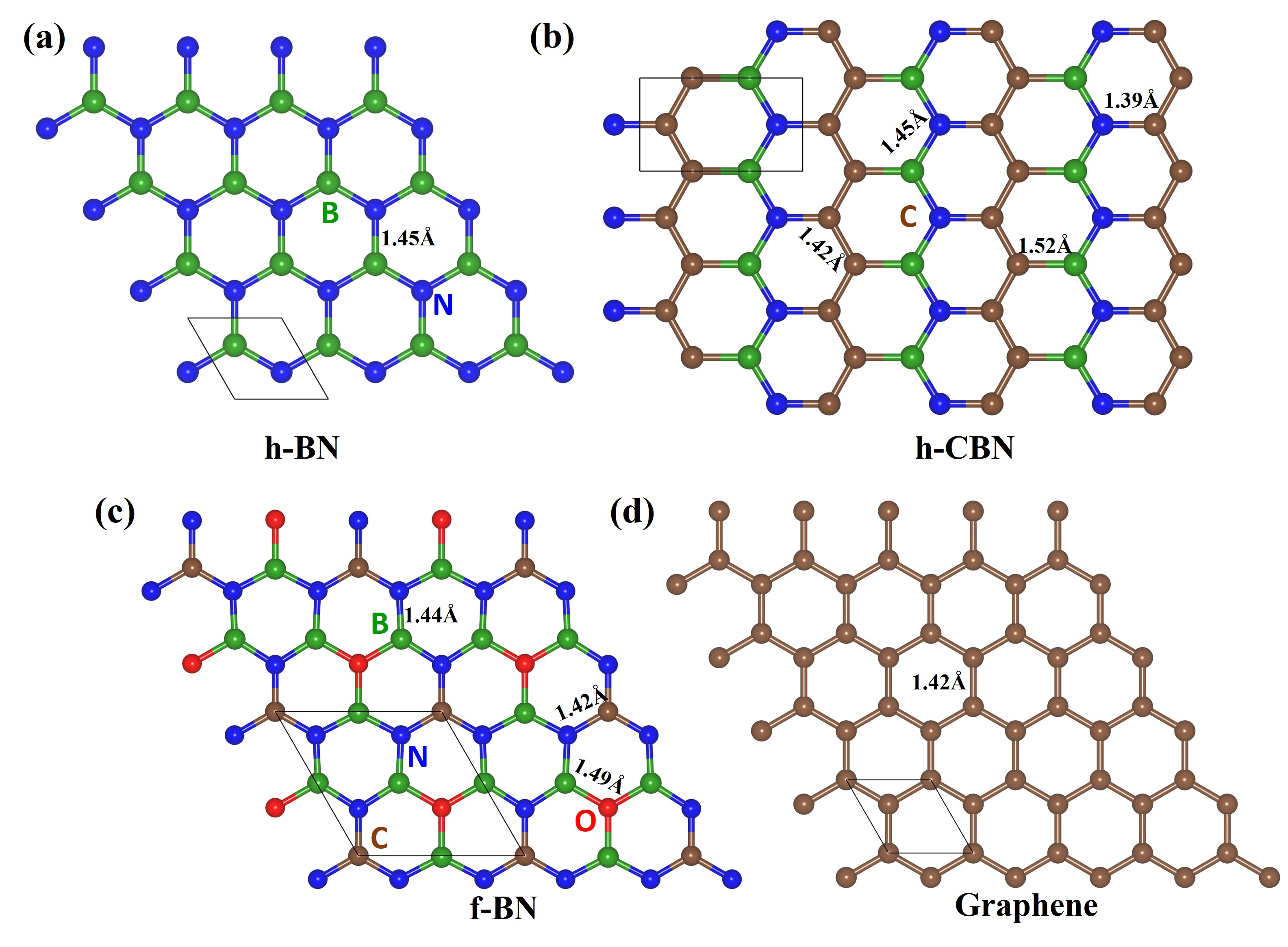}
    \caption{Optimized crystal structure of (a) h-BN, (b) h-CBN, (c) f-BN, and (d) graphene}
    \label{fig:fig_1}
\end{figure}

 A monolayer of 2D hexagonal boron nitride h-BN (Figure \ref{fig:fig_1}a) was considered with optimized in-plane lattice parameters a and b are 2.51\AA\ and the B-N bond length of 1.45\AA\ matches well with the earlier reports\cite{demiroglu2021extraordinary,sevik2014assessment,olaniyan2021first}. Figure \ref{fig:fig_1}b represents the alternating chains of carbon and boron nitride (h-CBN)\cite{sevik2014assessment,olaniyan2021first} with a rectangular lattice of the primitive cell containing two carbon atoms, one boron, and one nitrogen atom (C$_2$BN). The in-plane optimized lattice parameters are a=2.49\AA\ and b=4.35\AA\ which matches well with the earlier report\cite{olaniyan2021first}. In the primitive cell of h-CBN, it was observed that every C atom bonded with two other C atoms and either an N or B atom, while every N and (or B) atom is connected to two B (or N) atoms. The computed bond lengths of C-C, C-N, C-B, and B-N are 1.42, 1.39, 1.52, and 1.45\AA\ respectively\cite{olaniyan2021first}. For the functional boron nitride\cite{ren2021hydroxylated} (f-BN) the obtained optimized bonds are C-N, B-N, and B-O and their corresponding bond lengths are 1.42, 1.44, and 1.49\AA\ respectively (Figure \ref{fig:fig_1}c). Very small change in the B-N bond length is observed in h-CBN and f-BN (1.44\AA) corresponds to the pristine h-BN (1.45\AA) and this could impact the stability of the h-CBN and f-BN system\cite{olaniyan2021first}. Figure \ref{fig:fig_1}d represents optimized graphene structure with C-C bond length of 1.42\AA. 


\begin{figure}[htp]
    \centering
    \includegraphics[width=16.5cm]{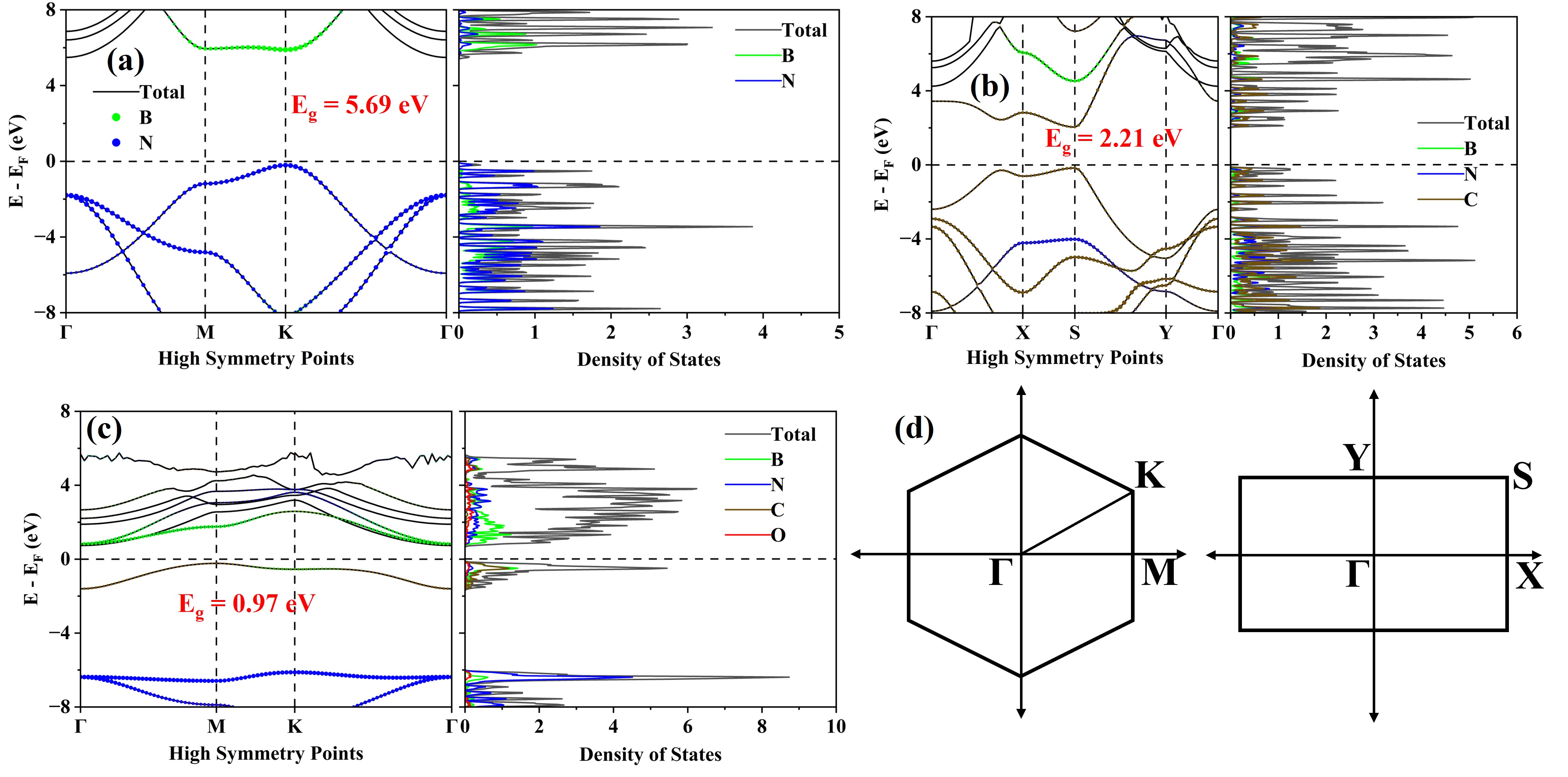}
    \caption{Bands and Density of States (DOS) along high symmetry points of (a) h-BN, (b) h-CBN, and (c) f-BN.}
    \label{fig:fig_2}
\end{figure}

The electronic density of states (DOS) and band structures along the high-symmetry points of the Brillouin zone reveal the electronic behavior of the materials. In this study, we compare the electronic properties of h-BN and h-CBN with those of f-BN. The highly stable h-BN exhibits a large semiconducting band gap of approximately 5.7eV \cite{pakdel2014nano}, which limits its applications in electronics and energy conversion \cite{fan2022predicting}. However, researchers have demonstrated that by tuning its electronic and structural properties, a variety of new applications can be realized \cite{cai2024boron, wang2022two}. Alongside h-BN, we have explored the projected band structures and DOS of h-CBN and f-BN, as shown in Figure \ref{fig:fig_2}.

Using the HSE06 functional, the calculated indirect band gap of monolayer h-BN is 5.69eV (Figure \ref{fig:fig_2}a), in good agreement with previous reports \cite{pakdel2014nano, pakdel2012facile, ci2010atomic}. The valence and conduction bands are predominantly composed of N and B 2$p$ orbitals, respectively. Upon transitioning to graphitic boron nitride (h-CBN), the band gap becomes direct and reduces significantly to 2.21eV (Figure \ref{fig:fig_2}b), with both the valence band maximum (VBM) and conduction band minimum (CBM) derived mainly from C 2$p$ orbitals at the same k-point. The formation of C–N and C–B bonds (Figure \ref{fig:fig_1}b) promotes $\pi$–$\pi$ delocalization among N 2$p_z$ electrons, contributing to the band gap reduction. This result is consistent with earlier theoretical studies \cite{olaniyan2021first, liu1989atomic, kaner1987boron, miyamoto1994chiral, ci2010atomic}. The narrowed band gap enhances the potential utility of these materials in electronic and energy-related applications \cite{pakdel2014nano}.

\begin{figure}[htp]
    \centering
    \includegraphics[width=14cm]{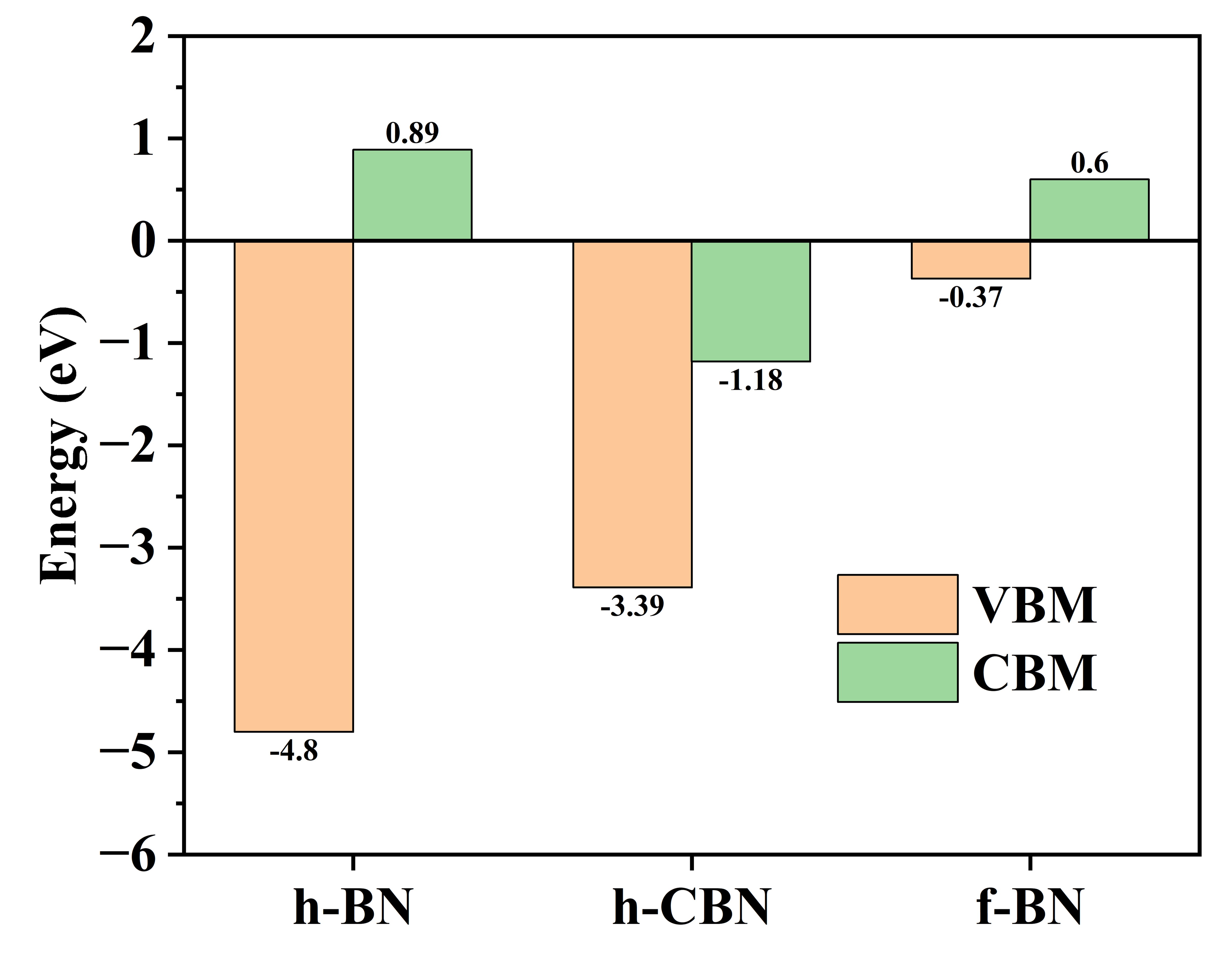}
    \caption{Comparision between the band edge position of all three systems  h-BN, h-CBN, and f-BN.}
    \label{fig:fig_3}
\end{figure}

C and O co-doping of h-BN further tailors its electronic structure, yielding a reduced indirect band gap of 0.97eV (Figure \ref{fig:fig_1}c). DOS analysis reveals substantial contributions from all atomic species to both the valence and conduction bands. A comparison of band edge positions (Figure \ref{fig:fig_3}) indicates a notable downward shift of both VBM and CBM for h-CBN and f-BN relative to pristine h-BN.

\begin{figure}[htp]
    \centering
    \includegraphics[width=15.5cm]{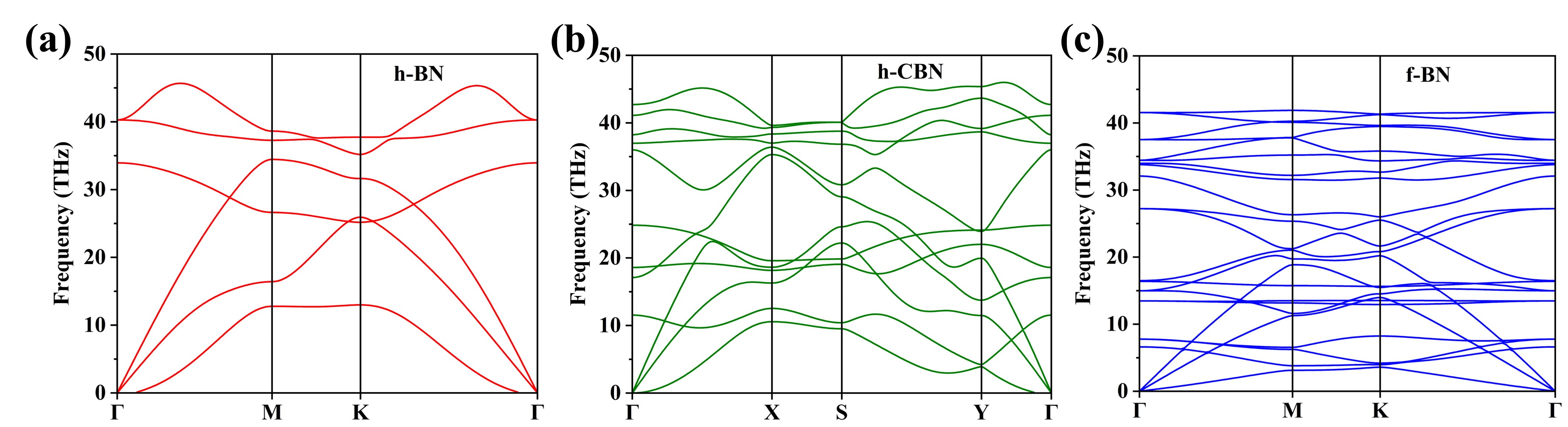}
    \caption{Phonon dispersion relation of (a) h-BN, (b) h-CBN, and f-BN.}
    \label{fig:fig_31}
\end{figure}

The dynamical stability of f-BN, along with h-BN and h-CBN, was assessed through the phonon dispersion calculations, as shown in Figure \ref{fig:fig_31}. The absence of imaginary frequencies across the entire Brillouin zone confirms the dynamical stability of all three systems. In each case, the vibrational spectra exhibit linear longitudinal and transverse acoustic (LA and TA) phonon modes, while the out-of-plane transverse acoustic (ZA) modes of h-BN and h-CBN display a characteristic quadratic dispersion near $q = 0$ (Figures \ref{fig:fig_31}a,b), consistent with earlier studies \cite{olaniyan2021first, sevik2014assessment}. In contrast, f-BN exhibits a linear dispersion for the ZA mode (Figure \ref{fig:fig_31}c), marking a distinct difference from h-BN and h-CBN. This altered behavior of the TA mode in f-BN is expected to have significant implications for its thermodynamic properties.

Thermal expansion study has been performed for this three system, during the growth of these materials on different substrates like silicon, silica / quartz (SiO$_2$) the thermal expansion of the lattice in the plane at the interface plays an important role. At the interface the in-plane lattice thermal expansion plays a significant role in controlling the many electronic transport phenomena. In our study of thermal expansion calculations, we did not consider any composite or interface structure, instead we focused on each bare system of h-BN, h-CBN, f-BN and quartz (SiO$_2$) to avoid the expensive computational cost. We also compared the thermal expansion of f-BN with that of graphene along with h-BN and hCBN. 

The coefficient of thermal expansion ($\alpha$) of these 2D materials is negative, with a temperature contraction or negative thermal expansion (NTE) occurring up to a certain range of temperatures. The origin of such exciting behavior can be explained both by theory and experiment\cite{liang2021negative}, most recent works have remarked that the interaction between phonons and anisotropic elasticity and acoustic phonons of a material can be important in understanding the origin of NTE behavior\cite{liang2021negative}. The effect of harmonic and anharmonic oscillators with interatomic distance plays a crucial role in the thermal expansion of all materials\cite{liang2021negative}. In particular, the higher-order term (anharmonic oscillator) of the interatomic potential manifests the NTE behavior. Many experimental and theoretical investigations demonstrate that the low-frequency phonon modes involving transverse and librational vibrations contribute to the NTE of framework structures of solids.   

\begin{figure}[htp]
    \centering
    \includegraphics[width=16cm]{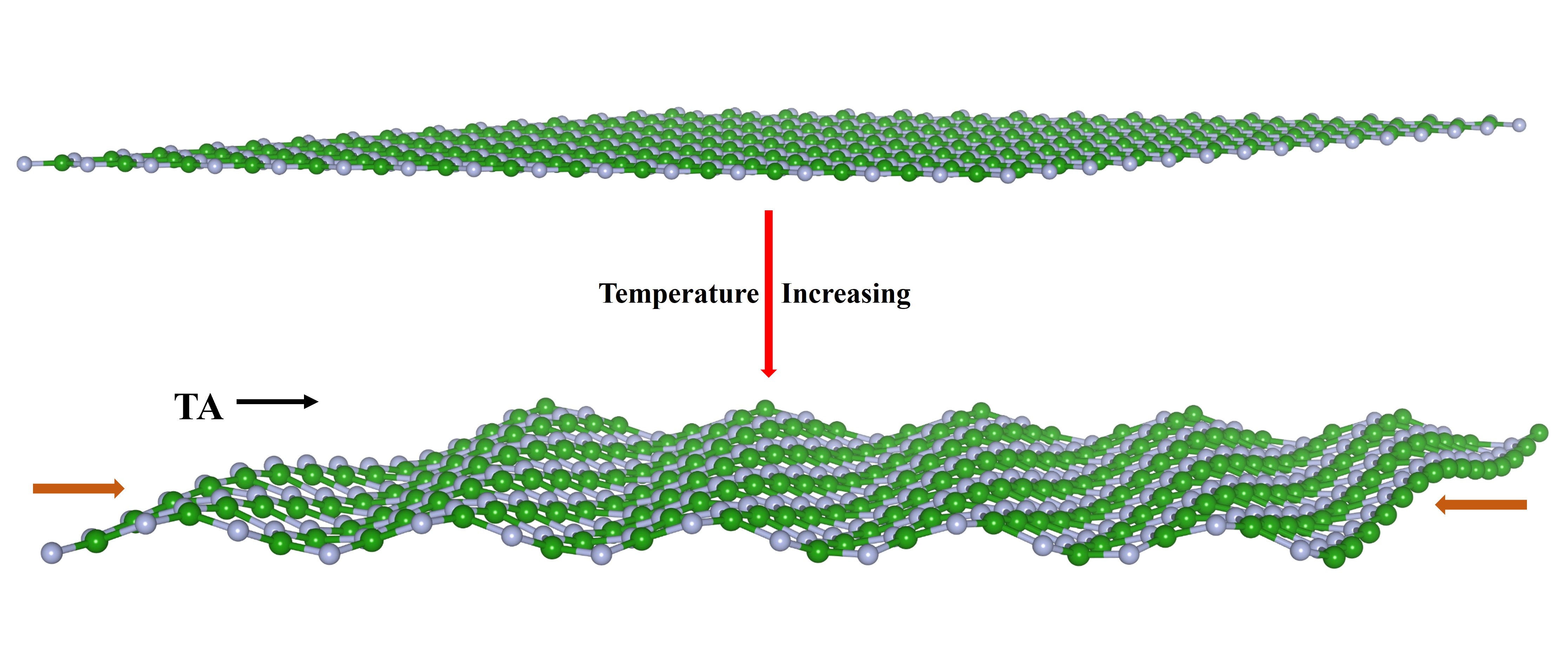}
    \caption{NTE in 2D single layer due to collective acoustic mode.}
    \label{fig:fig_5}
\end{figure}

The interplay between the transverse acoustic (TA) and longitudinal acoustic (LA) modes with temperature is activated and atomic vibrations are controlled by expansion or shrinkage\cite{liang2021negative}. In 2D material, collective out-of-plane vibrations (acoustic mode, or ZA mode) of atoms will cause a shrinkage (Figure \ref{fig:fig_5}) of the layer and dominate the origins of NTE in 2D materials\cite{liang2021negative}.

\begin{figure}[htp]
    \centering
    \includegraphics[width=16cm]{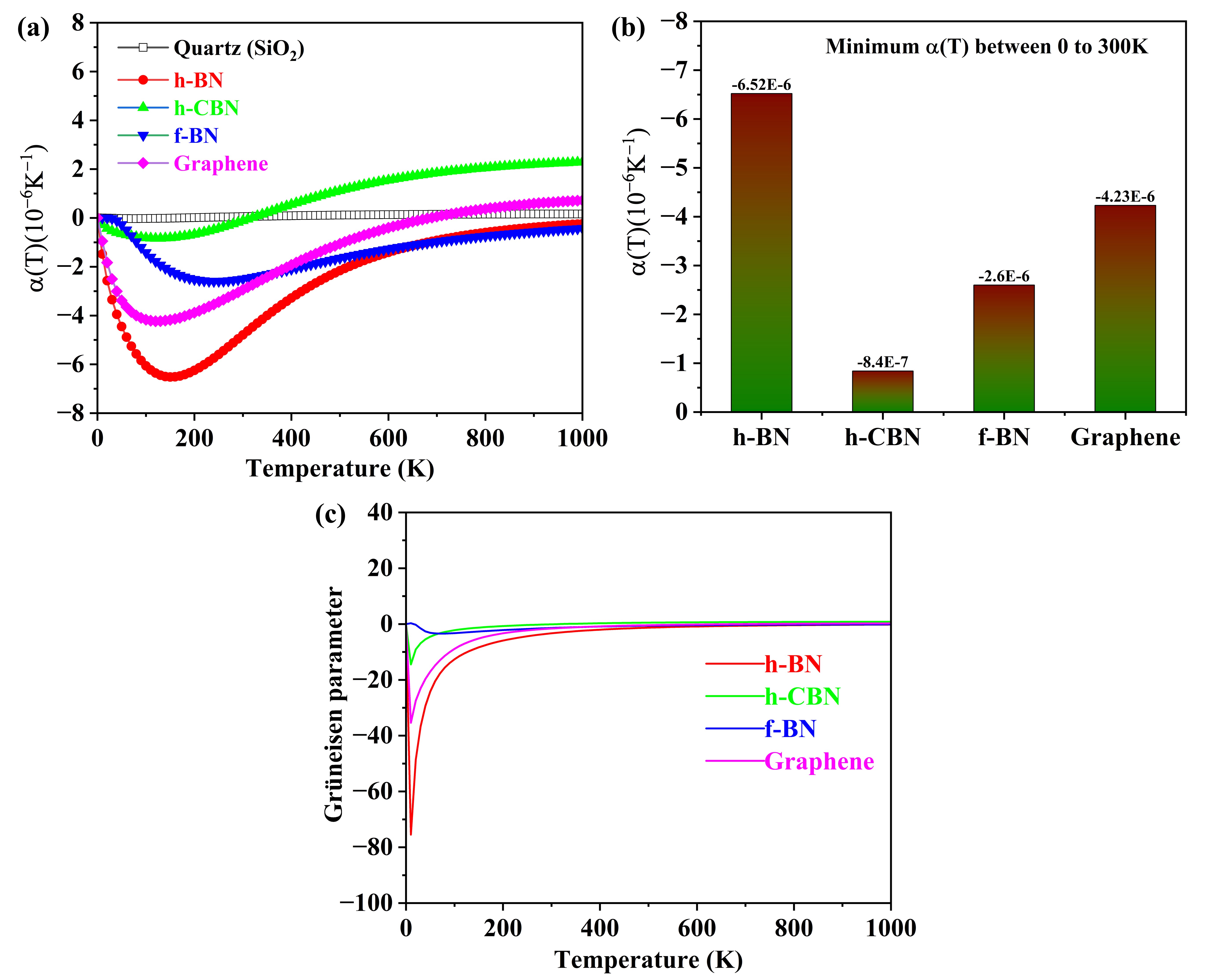}
    \caption{(a) Linear thermal expansion coefficient (LTEC) of h-BN, h-CBN, f-BN, and graphene (b) range of LTEC value of all these systems between 0 to 300K. (c) Gruneisen parameter.}
    \label{fig:fig_4}
\end{figure}

We first calculate the linear thermal expansion of quartz and it shows $\alpha$ very negligible with the temperature shown in Figure \ref{fig:fig_4}a (black curve) and agrees well with previous experimental and theoretical reports\cite{tian2016temperature,hahn1972thermal}.  Now, we have calculated the linear thermal expansion coefficient (LTEC) of the 2D monolayer h-BN, h-CBN, f-BN and graphene structure. Figure \ref{fig:fig_4} represents the LTEC plot for all systems, it is an important property of materials that is often considered in many technological applications\cite{drebushchak2020thermal}. The crystal shape and in-plane arrangement of the atoms for all BN base systems (Figure \ref{fig:fig_1}) affect the LTEC values; here our focus is especially on functional boron nitride or f-BN and we have designed f-BN with the stoichiometric composition of B$_{0.38}$N$_{0.38}$C$_{0.06}$O$_{0.18}$ 
Here we present our results on the LTEC of functional boron nitride (f-BN) using the QAH\cite{abraham2018thermal,togo2013phonopy,togo2015first,demiroglu2021extraordinary} approximation at the GGA level. f-BN was shown to comprise carbon, oxygen, and boron nitride chains (Figure \ref{fig:fig_1}c). Therefore, we first present the LTECs of pristine h-BN and hybrid h-CBN to validate our computational procedures by comparing our results to the thermodynamic properties of h-BN, h-CBN and graphene reported in other experimental\cite{slack1975thermal} and theoretical\cite{demiroglu2021extraordinary,wang2017mechanical,brito2020thermodynamic,sevik2014assessment,olaniyan2021first} studies. Furthermore, we compared the computed LTEC properties of h-BN, h-CBN, and graphene with those of f-BN to assess the uniqueness of the LTECs of all of the parent materials.  
We investigate the LTEC using DFPT and QHA by applying a considerably small lattice strain at ±0.5\% for 2D systems and we observed that our calculated LTECs (Figure \ref{fig:fig_4}a) of all the BN-based systems such as 2D monolayer h-BN (red), h-CBN (green), graphene (pink) and f-BN (blue) have negative LTECs of the same order of 10$^{-6}$ K$^{-1}$, which is more prominent at low temperatures between 0 to 300K. 

The negative LTECs or thermal contraction over a broad spectrum of temperatures is more pronounced in h-BN and f-BN than in h-CBN but the temperature dependence of the LTEC for all 2D materials shows the same trend; it has a minimum within the range of 0 to 300 K and it's shown in Figure \ref{fig:fig_4}b. The LTEC reaches a minimum of -6.5 x 10$^{-6}$ K$^{-1}$ for h-BN, -1.05 x 10$^{-6}$ K$^{-1}$ for h-CBN, -4.23 x 10$^{-6}$ K$^{-1}$ for graphene, and -2.6 x 10$^{-6}$ K$^{-1}$ for f-BN, then it rises monotonically with temperature. Since it is evident that by designing different f-BN materials could possibly alter the LTEC value and help to thermal expansion during the formation of these material on different substrate. Our findings demonstrate that the optimized composition achieves a 33.7
 

Anharmonicity in phonon modes is directly reflected by the Grüneisen parameter, which is also related to the third-order force constant \cite{wallace1972thermodynamics,togo2015first,villarreal2021metric}. Figure \ref{fig:fig_4}c shows the Grüneisen parameter for all systems, where the negative values between 0 and 500K suggest negative thermal expansion (NTE) behavior and signatures of transverse acoustic modes \cite{ritz2019thermal}. The introduction of C and O atoms into a h-BN sheet at B and N positions, forming f-BN, affects interatomic potentials and atomic bonds. This potentially impacts the transverse acoustic mode in a manner similar to pure h-BN and graphene, aiding in tuning their NTE behavior.

\section{Conclusions}
In this study, we have designed a new functionalized boron nitride (f-BN) system and systematically explored its structural, electronic, and thermodynamic properties, with particular emphasis on the coefficient of thermal expansion ($\alpha(T)$). In both f-BN and h-CBN, a noticeable change in the in-plane bond lengths relative to pure h-BN is observed, contributing to significant modifications in their electronic properties. The incorporation of C and O atoms, through their 2$p$ orbital contributions, leads to a substantial reduction in the electronic band gap down to 0.97 eV for f-BN and 2.21 eV for h-CBN compared to pristine h-BN. This band-gap engineering suggests promising potential for a wide range of applications.

Furthermore, the thermal expansion behavior reveals that at low temperatures, $\alpha(T)$ for f-BN and h-CBN is lower than that of h-BN. Notably, f-BN maintains a negative thermal expansion up to 1000 K, following a trend similar to h-BN, whereas h-CBN transitions to a positive $\alpha(T)$ at higher temperatures. This distinct thermal expansion behavior highlights the potential for precise control of thermal expansion mismatch with substrates like quartz, offering significant advantages for thermal management in device applications.



\begin{acknowledgement}
This work was supported by an NRF grant funded by MSIP, Korea (No. XXXXXXXX and No.
XXXXXXXXXXX), the Convergence Agenda Program (CAP) of the Korea Research Council of
Fundamental Science and Technology (KRCF) and the GKP project (Global Knowledge Platform) of the
Ministry of Science, ICT and Future Planning. The authors would also like to acknowledge support from the Korea Institute of Science and Technology Information (KISTI) Supercomputer Center through their RnD innovation support program (Grant No. KSC-2022-CRE-0510).
\end{acknowledgement}

\bibliography{achemso-demo}

\end{document}